\documentclass[11pt]{article}
\usepackage[a4paper, margin=1.3in]{geometry}

\usepackage{amsmath}
\usepackage{amssymb}
\usepackage{dsfont}
\usepackage{tikz}
\usetikzlibrary{bayesnet}
\usetikzlibrary{positioning,arrows.meta,shapes,calc}
\usepackage{makecell}
\usetikzlibrary{arrows}

\usepackage{authblk}
\usepackage[utf8]{inputenc}
\usepackage[numbers,sort]{natbib}
\usepackage{graphicx, url, algorithm2e}
\usepackage{centernot}
\newcommand{\indep}{\perp \!\!\! \perp}

\begin{document}

\title{Surjective Independence of Causal Influences for Local Bayesian Network Structures}

\author[1]{Kieran Drury\thanks{Corresponding author: kieran.drury@warwick.ac.uk}}
\author[1]{Martine J. Barons}
\author[1]{Jim Q. Smith}

\affil[1]{Department of Statistics, University of Warwick, Coventry, CV4 7AL, UK}


\date{29 September 2025}
\maketitle

\begin{abstract}
The very expressiveness of Bayesian networks can introduce fresh challenges due to the large number of relationships they often model. In many domains, it is thus often essential to supplement any available data with elicited expert judgements. This in turn leads to two key challenges: the cognitive burden of these judgements is often very high, and there are a very large number of judgements required to obtain a full probability model. We can mitigate both issues by introducing assumptions such as independence of causal influences (ICI) on the local structures throughout the network, restricting the parameter space of the model. However, the assumption of ICI is often unjustified and overly strong. In this paper, we introduce the surjective independence of causal influences (SICI) model which relaxes the ICI assumption and provides a more viable, practical alternative local structure model that facilitates efficient Bayesian network parameterisation.

\end{abstract}


\vspace{0.2cm}
\noindent\textbf{Classifications}: 62H22, 62C99, 68T30, 
68T37

\section{Introduction}
Bayesian network (BN) modelling \cite[see e.g.][]{Pearl1988ProbReasoning,JensenNielsen2007,FentonNeil2013Book,KorbNicholson2011} has now been a highly established, reliable and intuitive tool among statisticians, computer scientists and AI practitioners for a number of decades. One powerful use of BNs is as a decision support tool - modelling a complex real-world system to test potential policies before implementation by a decision centre \cite[see e.g.][]{JensenNielsen2007,KorbNicholson2011,Smithbook2010}. This use of BNs has become widespread in the 21st century due to the increased complexity and interconnectedness of the environments in which many decision problems are based.

One fundamental challenge - which we have experienced in many applications - of BN modelling, especially for decision support, is the lack of sufficient data for fully parameterising a model \citep{French2021, Werner2017}. BNs piece together sub-systems from a variety of distinct domains, leading to a large number of complex relationships with high-order interactions needing to be modelled (see examples in \cite{KorbNicholson2011,Barons2022foodsecurity}). We therefore often need to rely on expert judgement to parameterise each of these relationships \cite[see e.g.][]{Burgman2015,French2021,ohaganbook,Werner2017}.

Expert judgement elicitation does not, however, come without its own problems. If the system is too complex to be modelled through the available data, an insufficiently managed elicitation process can easily become intractable \citep{Werner2017}. This can happen in several ways - a lack of available, sufficiently knowledgeable experts; a lack of time and money available for eliciting this knowledge; the inability to suppress cognitive biases during the elicitation; and the difficulty of translating experience and knowledge into the required probabilistic assessments \citep{KorbNicholson2011,Woudenberg2015,Burgman2015,OHagan2019,ohaganbook}. These issues form the so-called ``knowledge bottleneck" \citep{KorbNicholson2011}.

Even if the above issues are addressed through careful structuring of the elicitation process, two significant problems persist. The first is simply the number of probabilistic judgements that are required to embellish the whole network \citep{Smithbook2010}. The second is that the judgements required from experts are often highly complex due to high-order interactions that are present in BNs, and because the judgements required from experts are usually probabilistic (see \cite{Woudenberg2015} for an example of this in practice). These issues lead to a high elicitation burden for the experts \citep{Werner2017}. They often become fatigued and more susceptible to a number of cognitive biases when this burden is high, threatening to corrupt the judgements they provide even further \citep{Burgman2015, Barons2022ElicitationBurden}. One way to reduce this elicitation burden is to restrict the model space through applying local structure assumptions across the network, often by applying particular causal interaction models on the local structures \citep{DiezDruzdzel2006}.

A popular class of causal interaction model relies on the assumption of \textit{independence of causal influence} (ICI) \cite{Heckerman1993,ZhangPoole1996} which assumes that each parent node influences the child node independently. ICI has been utilised within many sub-classes of local BN structure models such as noisy-OR \citep{Pearl1988ProbReasoning} and its extensions \citep{Diez1993,Henrion1989,Srinivas1993}, as well as in CPT interpolation methods such as those reviewed in \cite{Mkrtchyan2016,Blomaard2025}. Despite this, while it does simplify the elicitation process, its underlying assumptions are usually too strong and rigid to faithfully represent experts' beliefs about complex real-world systems.

In this paper, we present a simple, practicable methodology for modifying a network structure into a form that better incorporates the ICI assumption, while allowing more freedom for interactions between parents for whom the original ICI model would be too rigid. Akin to the ICI model, our recently developed local structure model - named the \textit{surjective independence of causal influences} (SICI) model - introduces a set of latent causal mechanisms acting as mediators between a child node and its parent set. Whereas ICI forces a bijection between the parent set and this mechanism set, SICI allows a more general surjective mapping between these sets, enabling the parents to be partitioned into blocks that themselves exhibit the ICI property. We can therefore modify existing CPT approximation methods for use within the SICI framework. The SICI model thereby allows quantitative embellishment of a BN to be performed with a significantly reduced burden in a way that is flexible enough to more faithfully model expert beliefs about the real-world system.

The paper is laid out as follows. In Section \ref{BBNs}, we review Bayesian networks and explore how they can be elicited through expert judgement. In Section \ref{ICI}, we explore the assumption of independence of causal influences and detail some of its uses. In Section \ref{SICI}, we introduce the surjective independence of causal influences model, detailing three specific variants of the model that allow for different combinations of stochastic and deterministic relationships between nodes. We detail their mathematical foundations and provide some practical examples of such models. We then explore how this modified network structure more flexibly accommodates the assumption of the ICI property, enabling efficient yet faithful elicitation of BN models. The paper concludes with a brief discussion and evaluation of the SICI methodology and future research directions.

\section{Bayesian Networks and their Elicitation} \label{BBNs}

Bayesian networks \cite[see e.g.][]{Pearl1988ProbReasoning,KorbNicholson2011,JensenNielsen2007,FentonNeil2013Book} are a type of probabilistic graphical model often used to model complex systems with many interdependencies through the graphical structure $G=(V,E)$. They are most often used in the discrete, static form in which each node is discrete and the model is run over just one time slice. Alternative forms of BN include hybrid Bayesian networks which permit continuous nodes \cite{Neil2008,FentonNeil2013Book} and dynamic Bayesian networks which include time-lag dependencies \cite{KorbNicholson2011,JensenNielsen2007}. This paper focuses on the parameterisation of discrete Bayesian networks - also applicable to dynamic Bayesian networks with discrete nodes. Henceforth we assume any Bayesian network to be discrete and static.

\vspace{0.5cm}
\begin{figure}[ht]
\centering
\scalebox{1.15}{
\begin{tikzpicture}
     \node[latent] (X1) {$X_1$};
     \node[latent,right=of X1] (X2) {$X_2$};
     \node[latent,right=of X2] (X3) {$X_3$};
     \node[latent,below=of X2,xshift=0cm] (X4) {$X_4$};
     \node[latent,right=of X4](X5){$X_5$};
     \node[latent] (X6) at ($ (X3)!0.5!(X5) + (1.5cm, 0) $) {$X_6$};
     \edge {X1,X3,X2} {X4};
     \edge {X3,X4} {X5};
     \edge {X3,X5} {X6};
\end{tikzpicture}
}
\caption{Example Bayesian network structure - a DAG on 6 nodes}
\label{fig:BN}
\end{figure}
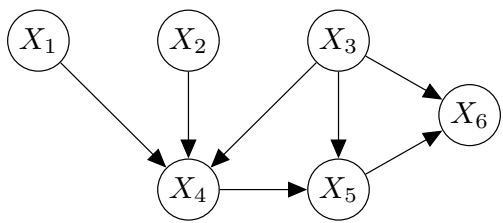
\vspace{0.25cm}
An example BN structure is shown in Figure \ref{fig:BN} with which we demonstrate some BN terminology. Each node, $X_i$, has a parent set, Pa$(X_i)=\{X_j:(X_j,X_i)\in E\}$, referring to the set of nodes from which you can reach $X_i$ in one step. Each $X_j$ would be called a parent of $X_i$, with $X_i$ referred to as their child. For example, Pa$(X_5)=\{X_3,X_4\}$. Nodes with an empty parent set - here being $X_1$, $X_2$ and $X_3$ - are called root nodes. Nodes with no children - here being just $X_6$ - are called leaf nodes, and are often a target variable of interest. We define grandparents, grandchildren etc. in similar ways. Any node $X_j$ which is a parent, grandparent, great-grandparent (and so on) of $X_i$ is called an ancestor of $X_i$, with $X_i$ being a descendent of $X_j$. Further BN terminology and examples can be seen in, for example, \cite{Pearl1988ProbReasoning,KorbNicholson2011,Smithbook2010,JensenNielsen2007}.

Bayesian networks are generally constructed using either data or elicited expert judgements. The methodology presented in this paper addresses the issue of parameterising a BN - a particularly challenging task due to the extreme number of parameters required throughout even moderately sized models. There can be significant data requirements for learning the quantitative parameters of a BN \cite{Ji2015}, and even more so for learning its structure \cite{Kitson2023}. Therefore, expert judgement is often required to construct a Bayesian network model. For some modelling problems, data is available but not to a sufficient quantity to produce reliable parameter estimates, in which case expert judgement can be used to complement the available data \cite{Zhou2014}. In some other cases, such little high-quality data is available for use that expert judgement must act as the primary source of information with which to build a model, possibly alongside secondary sources of information such as open-source literature on the domain. Below, we briefly describe the process of eliciting a Bayesian network through expert judgement.

There are two main stages to eliciting a Bayesian network. The first is the construction of the network structure - i.e. which variables to include, how to define them, the possible values they have and which variables to draw edges between. This is referred to as the \textit{qualitative stage} of the process, utilising \textit{qualitative} or \textit{soft elicitation} of expert judgements \cite{Cainsmethod,KorbNicholson2011,Wilkerson2021}. The second stage involves quantifying the relationships within the network. Modelling discrete BNs, as we assume in this paper, and as is common in practice, involves populating each non-root node's conditional probability table (CPT), as well as a marginal probability distribution for the root nodes. This is the \textit{quantitative stage} of the process, utilising \textit{quantitative elicitation} of expert judgements \cite{Cainsmethod,KorbNicholson2011,ohaganbook}.

The qualitative elicitation process typically consists of natural language discussions with domain experts in order to understand how \textit{they} picture the structure of the real-world system \cite{barons2018epj,Smithbook2010}. In contrast, the quantitative elicitation process can involve a high number of probabilistic judgements about high-order interactions that those not trained in probability struggle to instinctively comprehend. This renders the quantitative elicitation process the most burdensome stage for the experts. The number of probabilistic judgements required to populate each CPT is one part of this problem. Consider the child node $Y$ with parent nodes $\mathbf{X}=\{X_1,\ldots,X_n\}$. Let $s_1,\ldots,s_n$ denote the number of possible states for each of the parent nodes, and $s_c$ that for the child. The number of probabilities required to fully parameterise the CPT of $Y|\mathbf{X}$, accounting for each row of the CPT summing to one, is given by:
\begin{equation}\label{eq:CPTentries}
   N_{Y}= \left(\prod_{i=1}^ns_i\right)\cdot(s_c-1)
\end{equation}
The number of probabilistic judgements required across the network can quickly become enormous. This, combined with the high cognitive burden that probabilistic judgements about high-order interactions bring, makes direct quantitative elicitation often intractable.

The question of how to reduce the elicitation burden faced by experts is therefore of high importance. Methods for structuring the quantitative elicitation process such as the Delphi method \citep{Rowe1999}, the Sheffield Elicitation Framework (SHELF) \cite{Gosling2018} and the IDEA protocol \citep{Hanea2017IDEA} mitigate the effects of cognitive biases when providing probabilistic judgements, somewhat reducing the cognitive burden faced by experts. However, these methods do not reduce the number of judgements required, nor do they remove the probabilistic nature of these judgements, or the high-order interactions that these judgements condition on. There is therefore a practical need to reduce the elicitation burden further than what these methods provide. The main question is how to do this while maintaining faithfulness of the model to expert beliefs about the real-world system.

Consequently, methods have been developed for simplifying the structures found within BNs to reduce the number of quantitative assessments required to fully embellish the model. Variables modelled in a BN are often influenced by just a small subset of the other variables in the network through mechanisms that are invariant to variables outside this local structure \citep{Pearl2009}. The local structure we refer to in this paper simply considers a node $Y$ and its parent set pa$(Y)$. This local structure can be modified without impacting other local structures in the network due to the highly compartmentalised structure a BN exhibits. In this way, these local structures can be simplified through some assumption about how the causal mechanisms operate between a child and its parent set.

A number of local structure models have been developed that ease the quantitative elicitation process by embedding some such assumption. Many of these models require all nodes to be binary, including noisy-OR \citep{Pearl1988ProbReasoning} and its extensions \citep{Henrion1989,LemmerGossink2004,Quintanar-GagoNelson2021}, as well as the intercausal cancellation model \citep{Woudenberg2015}. A notable example that allows for $n$-ary nodes is the noisy-MAX model \citep{Diez1993,Srinivas1993}. Further details and extensions of these models are summarised in a previous review \cite{DiezDruzdzel2006}. Alternative, purely quantitative methods have been developed to interpolate or otherwise approximate missing CPT values, often just from assessments of the influence of each parent. Some such methods are analysed in previous reviews \cite{Mkrtchyan2016,Blomaard2025}, though further, recent CPT approximation methods exist beyond the scope of these reviews \cite{Hassall2019,Phillipson2021,Mascaro2022}. The above methods each restrict the parameter space of a particular CPT, enabling the CPT to be approximated through a much smaller number of expert judgements than direct elicitation would require.

These methods each construct approximate CPTs through measures of influence of each individual parent, or through interpolation between elicited rows of the CPT considering one parent change at a time. The modelled information therefore often solely concerns the influence of each individual parent, ignoring interactions between parents. Some approximate conditional probability mass function that combines these marginal contributions without embedding any conditioning interaction terms is used to model the child node. Such an approximation would be valid if the influence of each parent on the child node was independent of the values taken by the other parents - an assumption known as \textit{independence of causal influences}. In the next section, we explore this key assumption so that we can develop methodology addressing the representation of local network structures that justifies the use of models that utilise this property.

\section{Independence of Causal Influences} \label{ICI}
Though models utilising independence of causal influence (ICI) had previously been used implicitly \citep[e.g.][]{Pearl1988ProbReasoning}, the concept was first formally introduced under the name `causal independence' \cite{Heckerman1993}. ICI is a local structure assumption that simplifies the embellishment of a BN, tackling the challenges of BN parameterisation and quantitative elicitation discussed in Section \ref{BBNs}. While it can be a very strong assumption for some applications, it is a highly practicable methodology that provides significant parameter savings when parameterising a BN.

\begin{figure}[ht]
\centering
\scalebox{1.25}{
\begin{tikzpicture}
     \node[latent] (X1) {$X_1$};
     \node[latent,right=of X1,xshift=-0.8cm] (X2) {$X_2$};
     \node[latent,right=of X2,xshift=-0.8cm] (X3) {$X_3$};
     \node[latent,right=of X3,xshift=-0.8cm] (X4) {$X_4$};
     \node[latent,right=of X4,xshift=0.2cm] (Xn) {$X_n$};
     \node[latent,below=of X3,xshift=0cm] (Y) {$Y$};%
     \path (X4) -- node[auto=false]{\ldots} (Xn);
     \edge {X1,X2,X3,X4,Xn} {Y};
\end{tikzpicture}
}
\caption{Initial local structure consisting of child node $Y$ and its parent set $\mathbf{X}$}
\label{LocalStructure}
\end{figure}
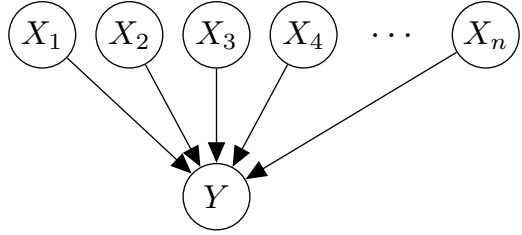

Consider a BN whose structure has already been elicited (or learnt), and denote the child node of a local structure by $Y$. Let its parents be written as Pa$(Y)=\mathbf{X}=\{X_1,\ldots,X_n\}$. We assume that no parents are adjacent for simplicity, though this need not be the case. The initial, unmodified local structure for the child node $Y$ is given in Figure \ref{LocalStructure}.

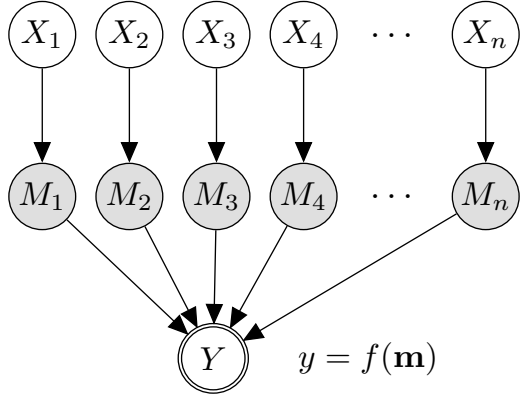
\begin{figure}[h]
\centering
\scalebox{1.25}{
\begin{tikzpicture}
     \node[latent,xshift=-1.2cm,style = double] (Y) {$Y$};%
     \node[obs,above=of Y,xshift=-1.8cm] (M1) {$M_{1}$}; %
     \node[obs,right=of M1,xshift=-0.8cm] (M2) {$M_{2}$}; %
     \node[obs,right=of M2,xshift=-0.8cm] (M3) {$M_{3}$}; %
     \node[obs,right=of M3,xshift=-0.8cm] (M4) {$M_{4}$}; %
     \node[obs,right=of M4,xshift=0.2cm] (Mn) {$M_n$}; %
     \node[latent,above=of M1] (X1) {$X_1$};
     \node[latent,right=of X1,xshift=-0.8cm] (X2) {$X_2$};
     \node[latent,right=of X2,xshift=-0.8cm] (X3) {$X_3$};
     \node[latent,right=of X3,xshift=-0.8cm] (X4) {$X_4$};
     \node[latent,right=of X4,xshift=0.2cm] (Xn) {$X_n$};
     \node[right=of Y,yshift=0cm,xshift=-0.6cm] (f){\small{$y=f(\mathbf{m})$}};;
     \path (M4) -- node[auto=false]{\ldots} (Mn);
     \path (X4) -- node[auto=false]{\ldots} (Xn);
    \coordinate[right of=X4,xshift=-0.35cm,yshift=-0.2cm] (d1);
    \coordinate[right of=d1,xshift=-0.4cm] (d2);
     \edge {M1,M2,M3,M4,Mn} {Y};
     \edge {X1} {M1};
     \edge {X2} {M2};
     \edge {X3} {M3};
     \edge {X4} {M4};
     \edge {Xn} {Mn};
\end{tikzpicture}
}
\caption{Local ICI model structure}
\label{LocalICIStructure}
\end{figure}

Now suppose we know, through expert judgement or otherwise, that each parent independently influences the value taken by the child, and thus this local structure satisfies the ICI property across its parent set. This can be thought of as each parent influencing the child node through its own independent causal mechanism. We can therefore modify our representation of the local structure by introducing a set of mechanisms, one for each parent, that explicitly quantifies the individual effect of each parent on the child \citep[see e.g.][]{Heckerman1993,HeckermanBreese1996,vanGerven2008}. Essentially, this uses a holistic approach to node divorcing \cite[see e.g.][]{Olesen1989,Rohrbein2009,Rosenkrantz2025,Cainsmethod}, requiring that every parent be divorced from the child. The approach here is not as general as that seen by the broad class of causal interaction models \cite{MeekHeckerman1997} as the ICI model requires each parent to be divorced from the child through its own unique intermediate node. We denote these mechanisms by $\mathbf{M}=\{M_1,\ldots,M_n\}$ where each mechanism typically has the same set of potential values as $Y$. The ICI model structure is shown in Figure \ref{LocalICIStructure}, where the mechanism nodes are shaded grey as they do not necessarily explicitly represent variables in the real-world system, and where the child node is drawn with two concentric circles as it is deterministic.

\vspace{0.5cm}

The structure of the ICI model is defined by the parent set pa$(Y)=\mathbf{X}$, the corresponding mechanism set $\mathbf{M}$, and the following conditional independence statements that can be verified through d-separation \cite{Pearl1988ProbReasoning,lauritzen1996graphical} (where $V$ denotes all nodes across the whole network beyond the local structure in focus):
\vspace{-0cm}
\begin{align}
   & \hspace{-2cm} Y\indep \mathbf{X}\mid \mathbf{M} \label{eq:ICIindep1} \\ 
   & \hspace{-2cm} M_i\indep M_j\mid \mathbf{X} \label{eq:ICIindep2} \\
   & \hspace{-2cm} \mathbf{M}\indep \left(V\setminus\{\mathbf{X}\cup Y\}\right)\mid \{\mathbf{X}\cup Y\}\label{eq:ICIindep3} \\
   & \hspace{-2cm} M_i\indep\left( \mathbf{X}\setminus \left\{X_{i}\right\}\right)\mid X_{i}.\label{eq:ICIindep4}
\end{align}
Statement \ref{eq:ICIindep1} is required to ensure that the mechanism nodes fully quantify the parents' effects on the child. Statement \ref{eq:ICIindep2} enforces the underlying ICI assumption that the mechanism nodes are independent of one another given the parent nodes, conditioning on $\mathbf{X}$ to ensure the statement holds true even when some parents are directly connected to each other by an edge. Statement \ref{eq:ICIindep3} ensures that the introduction of the mechanism nodes - which does not introduce new information to the model but does facilitate more efficient CPT parameterisation - does not influence the values taken by nodes elsewhere in the network (i.e. not shown in Figure \ref{LocalICIStructure}). Finally, statement \ref{eq:ICIindep4} enforces the bijective requirement of the ICI model - that each mechanism has exactly one parent.

As a result of the above conditional independence statements, the ICI model structure is automatically defined once the parent set of $Y$ has been decided. Once this structure is decided, the task of parameterising the local structure can begin.


The ICI model is defined to have stochastic upper (parent-to-mechanism) relationships, while the lower (mechanism-to-child) relationship is modelled deterministically through the deterministic function $f$. Effectively, each mechanism node is modelled through a probability mass function reflecting the strength of support for each of the child node's states given the value taken by its parent. Then $Y|\mathbf{M}$ is modelled deterministically through the function $f$ that combines the values taken across the mechanism set $\mathbf{m}$. The CPT of $Y|\mathbf{X}$ is approximated under the ICI model through the following result \citep{vanGerven2008}, utilising the law of total probability and conditional independence statements \ref{eq:ICIindep1}-\ref{eq:ICIindep4}:

\vspace{-0.25cm}\begin{align}\label{ICIdef}
p(y|\mathbf{x})=&\sum_\mathbf{m}p(y|\mathbf{m},\mathbf{x})p(\mathbf{m}|\mathbf{x})=\sum_\mathbf{m}p(y|\mathbf{m})p(\mathbf{m}|\mathbf{x})=\sum_{\mathbf{m}|f(\mathbf{m})=y}p(\mathbf{m}|\mathbf{x}) \nonumber \\
=&\sum_{\mathbf{m}|f(\mathbf{m})=y}\;\prod_{i=1}^{n}p(m_{i}|x_i).
\end{align}

One particular generalisation of the ICI model permits a stochastic relationship between the mechanism nodes and the child node (i.e. the lower relationship), and is known as the \textit{probabilistic independence of causal influences} (PICI) model \cite{DiezDruzdzel2006,ZagoreckiDruzdzel2006,Zagorecki2006}. The PICI model has the same structure as the ICI model, and therefore also shares conditional independence statements \ref{eq:ICIindep1}-\ref{eq:ICIindep4}. However, rather than using a deterministic function to model the relationship $Y|\mathbf{M}$, we use the function $f:Y\times\mathbf{M}\rightarrow[0,1]$ to define $p(y|\mathbf{m})=f(y,\mathbf{m})$. We therefore obtain the following, more general result for approximating the CPT of $Y|\mathbf{X}$ under the PICI model \cite{DiezDruzdzel2006}:
\begin{align}
\hspace{-0.0cm}p(y|\mathbf{x})=\sum_\mathbf{m}p(y|\mathbf{m},\mathbf{x})p(\mathbf{m}|\mathbf{x})=&\sum_\mathbf{m}\left[p(y|\mathbf{m})\prod_{i=1}^np(m_i|x_i)\right] \nonumber \\
    =&\sum_\mathbf{m}\left[f(y,\mathbf{m})\prod_{i=1}^np(m_i|x_i)\right].\label{eq:PICI}
\end{align}

One example of a PICI model is the `PICI average model' \cite{ZagoreckiDruzdzel2006,DiezDruzdzel2006,Zagorecki2006} which uses the function:
\begin{equation*}
    \hspace{-0.75cm}f(y,\mathbf{m})=p(y|\mathbf{m})=\frac{1}{n}\left|\{m_i:m_i=y\} \right|=\frac{1}{n}\sum_{1=1}^n\mathds{1}_{\left\{m_i=y\right\}}
\end{equation*}
to model the stochastic relationship between the mechanisms and the child. The PICI average model has been utilised and evaluated within the modelling application of menstrual cycle predictions - a domain that suffers from data scarcity issues as discussed in this paper \cite{Zagorecki2016}.

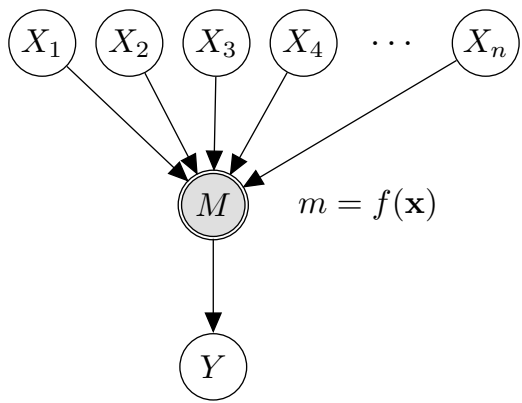
\begin{figure}[h!]
\centering
\scalebox{1.25}{
\begin{tikzpicture}
     \node[latent,xshift=-1.2cm] (Y) {$Y$};%
     \node[obs,above=of Y,style=double] (M) {$M$}; %
     \node[latent,above=of M1] (X1) {$X_1$};
     \node[latent,right=of X1,xshift=-0.8cm] (X2) {$X_2$};
     \node[latent,right=of X2,xshift=-0.8cm] (X3) {$X_3$};
     \node[latent,right=of X3,xshift=-0.8cm] (X4) {$X_4$};
     \node[latent,right=of X4,xshift=0.2cm] (Xn) {$X_n$};
     \node[right=of M,yshift=0cm,xshift=-0.6cm] (fm){\small{$m=f(\mathbf{x})$}};;
     \path (X4) -- node[auto=false]{\ldots} (Xn);
    \coordinate[right of=X4,xshift=-0.35cm,yshift=-0.2cm] (d1);
    \coordinate[right of=d1,xshift=-0.4cm] (d2);
     \edge {M} {Y};
     \edge {X1,X2,X3,X4,Xn} {M};
\end{tikzpicture}
}
\caption{Local SCM structure}
\label{LocalSCMStructure}
\end{figure}

So far, we have seen the ICI model that features stochasticity in its upper relationships and the PICI model that features stochasticity in the upper and lower relationships. It does not make reasonable sense to make the upper relationships deterministic due to each mechanism having only one parent. One simpler class of models that features deterministic relationships between the parent set and an intermediate layer is that of the \textit{simple canonical model} (SCM) \cite{DiezDruzdzel2006}. In this model, there is only one intermediate node, $M$. The parents $\mathbf{X}$ combine into $M$ through the deterministic combination function $f$ before the child is simply modelled through a conditional probability mass function $p(y|M)$ - conditional on just the one node. The SCM structure is shown in Figure \ref{LocalSCMStructure}. While the SCM is far from an ICI model, we introduce it here as it forms a class of models that can be used in applied BN modelling to facilitate efficient model parameterisation. It is by far the easiest model class to parameterise, requiring just two parameters if $M$ and $Y$ are binary, though it is far too simple a model to be informative and reliable in most cases. The surjective independence of causal influences model that we introduce in Section \ref{SICI} has similarities with not only the ICI model, but also with SCMs, hence we briefly present SCMs here as a point of comparison. 

\vspace{0.5cm}

Further subclasses of ICI models, namely amechanistic, decomposable, multiply decomposable and temporal ICI are described in \cite{HeckermanBreese1996}. These are beyond the scope of this introductory paper, but such properties will be evaluated in the future for the surjective independence of causal influences model, introduced in Section \ref{SICI}.

ICI models are useful when data is insufficient for providing reliable parameter estimates when parameterising the local structure of a BN. The key benefit of the ICI model is that it significantly reduces the number of parameters required for full parameterisation; it reduces the parameter growth in the number of parents, $n$, from exponential (for the original local structure - see Equation \ref{eq:CPTentries}) to linear in most cases \cite{vanGerven2008}. This significantly reduces the size of the parameter space of a CPT being modelled, thus reducing the data requirements for parameter learning algorithms or the resource requirements for performing a structured expert judgement elicitation. Hence, if the quantity of available data is limited (but is otherwise of good quality), transforming the local structure into an ICI model may be sufficient for yielding reliable parameter estimates that were previously unavailable. If the data is still insufficient for this, the ICI model facilitates far quicker elicitation of these parameters than eliciting parameters for the full CPT parameter space. Furthermore, it is important to note that eliciting experts' probabilistic judgements is highly challenging; domain experts often do not find this process intuitive and they struggle to process judgements when conditioning on multiple parents - as is almost always required for eliciting full CPTs. Parameterising a CPT within the ICI model simplifies this process not only through reducing the number of probabilistic judgements required from experts, but by reducing the typical number of conditioning variables in each probability assessment. In either case, the ICI model can bring with it notable practical benefits when it comes to model parameterisation. 

\begin{figure}[ht]
    \centering
    \scalebox{1.25}{
    \begin{tikzpicture}
     \node[latent,style = double] (Y) {$Y$};%
     \node[obs,above=of Y,xshift=-2.8cm] (M1) {$M_{1}$}; %
     \node[obs,above=of Y,xshift=0cm] (M2) {$M_{2}$}; %
     \node[obs,above=of Y,xshift=2.8cm] (Mm) {$M_{3}$}; %
     \node[latent,above=of M1,xshift=-0.6cm] (X1) {$X_1$};
     \node[latent,above=of M1,xshift=0.6cm] (I1) {$I_1$};
     \node[latent,above=of M2,xshift=-0.6cm] (X4) {$X_2$};
     
     \node[latent,above=of M2,xshift=0.6cm] (I2) {$I_2$};
     \node[latent,above=of Mm,xshift=-0.6cm] (Xn) {$X_3$};
     \node[latent,above=of Mm,xshift=0.6cm] (I3) {$I_3$};
     \node[below=of X2,yshift=0.4cm,xshift=-1.7cm] (f1){\small{$X_1\land\lnot I_1$}};
     \node[below=of X4,yshift=0.4cm,xshift=-0.45cm] (f2){\small{$X_{2}\land\lnot I_2$}};
     \node[below=of Xn,yshift=0.4cm,xshift=-0.45cm] (fm){\small{$X_{3}\land\lnot I_3$}};
    \coordinate[right of=X4,xshift=-0.35cm,yshift=-0.2cm] (d1);
    \coordinate[right of=d1,xshift=-0.4cm] (d2);
     \edge {M1,M2,Mm} {Y};
     \edge {X1,I1} {M1};
     \edge{X4,I2} {M2};
     \edge {Xn,I3}{Mm};
     \node[right=of Y,yshift=0cm,xshift=-1cm] (f){\small{$y=f(\mathbf{m})=\bigvee\limits_{i=1}^{3}M_i$}};;
    \end{tikzpicture}
    }
\caption{The noisy-OR model with three parents}
\label{NoisyOR1}
\end{figure}
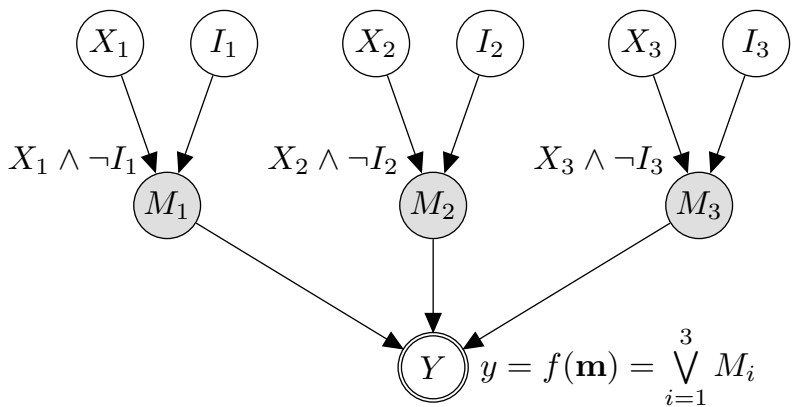

Some of the models described in Section \ref{BBNs} can easily be written as ICI models. For example, consider the noisy-OR model \citep{Pearl1988ProbReasoning} over a set of binary nodes. This model introduces inhibitor nodes $\mathbf{I}=\{I_1,\ldots,I_n\}$ that, combined with the parent set $\mathbf{X}$, define each of the mechanisms as presented here; $M_i=1$ (or `true' etc.) if the cause is present and the inhibitor node is false, else it takes value $0$ (i.e. $M_i=1\iff X_i=1 \;\land\; I_i=0$). Then the function $f$ is simply the deterministic OR function; if any of the mechanisms $M_i$ take the value $1$, the effect will be present (i.e. $Y=1\iff  M_1=1 \;\lor\; \ldots \;\lor\; M_n=1$) \citep{HeckermanBreese1996}. The noisy-OR model as described here is shown in Figure \ref{NoisyOR1}. 

\vspace{0.5cm}
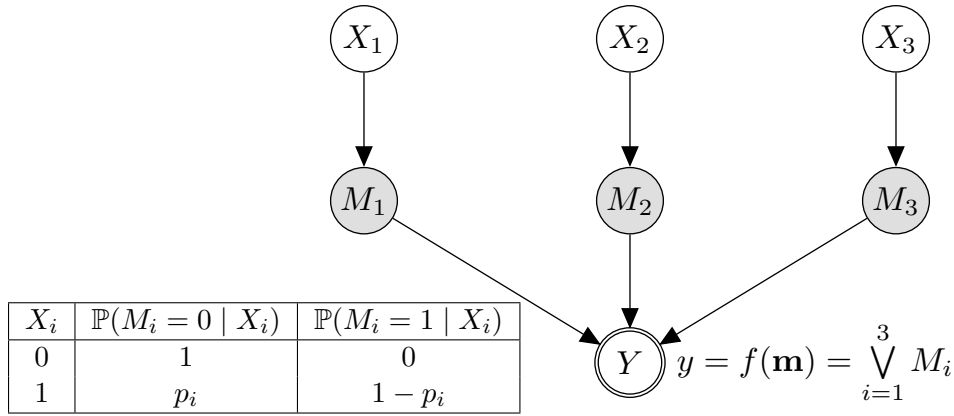
\begin{figure}[ht]
    \centering
    \scalebox{1.25}{
    \begin{tikzpicture}
     \node[latent,style = double] (Y) {$Y$};%
     \node[obs,above=of Y,xshift=-2.8cm] (M1) {$M_{1}$}; %
     \node[obs,above=of Y,xshift=0cm] (M2) {$M_{2}$}; %
     \node[obs,above=of Y,xshift=2.8cm] (Mm) {$M_{3}$}; %
     \node[latent,above=of M1] (X1) {$X_1$};
     \node[latent,above=of M2] (X4) {$X_2$};
     \node[latent,above=of Mm] (Xn) {$X_3$};
     \node[below=of X4,yshift=-1.3cm,xshift=-3.85cm] (f1){
\scalebox{0.8}{
\begin{tabular}{|c|c|c|}
\hline
$X_i$ & $\mathbb{P}(M_i=0\mid X_i)$ & $\mathbb{P}(M_i=1\mid X_i)$\\
\hline
$0$   &  $1$    &   $0$    \\
$1$   &  $p_i$    &   $1-p_i$    \\
\hline  
\end{tabular}
}
};
    \coordinate[right of=X4,xshift=-0.35cm,yshift=-0.2cm] (d1);
    \coordinate[right of=d1,xshift=-0.4cm] (d2);
     \edge {M1,M2,Mm} {Y};
     \edge {X1} {M1};
     \edge{X4} {M2};
     \edge {Xn}{Mm};
     \node[right=of Y,yshift=0cm,xshift=-1cm] (f){\small{$y=f(\mathbf{m})=\bigvee\limits_{i=1}^{3}M_i$}};
\end{tikzpicture}
}
\caption{The noisy-OR model as an explicit ICI model, featuring a general CPT for the mechanism node $M_i$}
\label{ICINoisyOR}
\end{figure}

We can equivalently present the noisy-OR model without the explicit use of the inhibitor nodes $\mathbf{I}$. In this case, we simply embed the probability $\mathbb{P}(I_i=1)=p_i$ into the CPT of $M_i|X_i$ for the row representing $X_i=1$. This equivalent representation of the noisy-OR model is shown in Figure \ref{ICINoisyOR}, demonstrating explicitly that the noisy-OR model is an ICI model as described in this paper.

A similar construction is simple for the noisy-MAX model \citep{Diez1993,Srinivas1993} which models $n$-ary variables with the combination function $f$ being the deterministic MAX function \citep{HeckermanBreese1996}. Many other models and methods given in Section \ref{BBNs} implicitly assume the ICI property, assuming that the probability mass function of the child can be approximated through a combination of the contributions made individually by each parent.

In the ICI model, an important underlying assumption is that there exists a bijection between the parents and the mechanisms. This comes from the idea that each modelled parent brings its own unique effect upon the child, and these effects are independent of each other. In practice, this would require that each parent in the real-world system affects the child through a mechanism unique to that variable, and that these mechanisms operate completely independently of each other. We argue that this is an excessively strong assumption, rendering the ICI model too restrictive as it does not allow any interactions between parents' causal mechanisms. For this reason, we have developed a generalisation of the ICI model that permits interactions between the causal mechanisms of particular subsets of parents. In the next section, we introduce this generalisation known as the \textit{surjective independence of causal influences} (SICI) model.

\section{The Surjective ICI Model} \label{SICI}
The surjective independence of causal influences (SICI) model generalises the ICI model by allowing the mapping $\phi:\mathbf{X}\rightarrow\mathbf{M}$ between the parent set and the mechanism set to be a surjection rather than a bijection. This allows multiple parents to feed into one shared mechanism, enabling the modelling of interactions between the causal mechanisms of particular parents. This is performed by partitioning the parent set $\mathbf{X}$ such that parents partitioned into the same block, $B_i$, feed into a common mechanism node, $M_i$. Parents are placed in the same block as others when the assumption of ICI breaks down between these parents. The SICI model therefore features $m\leq n=|\mathbf{X}|$ mechanism nodes, and we can more justifiably assume the ICI property to hold across these mechanism nodes $\mathbf{M}$ rather than across the original parents $\mathbf{X}$. 

The SICI model further generalises the ICI model by allowing stochasticity not only to be modelled between the parent nodes and the mechanism nodes, but also between the mechanism nodes and the child. We present three variants of the SICI model that model either the upper (parent-to-mechanism) relationships, the lower (mechanism-to-child) relationship, or all relationships stochastically.

No matter where the stochasticity in the local system is inputted, the SICI model structure remains the same. The structure of the SICI model is defined by the parent set pa$(Y)=\mathbf{X}$, the mechanism set $\mathbf{M}$ and the surjection $\phi:\mathbf{X}\rightarrow\mathbf{M}$ - which determines the edge set of the subgraph on $\mathbf{X} \cup \mathbf{M}$. As we assume that the structure of the original BN is already known through structure learning or expert elicitation, we already have the fixed parent set pa$(Y)=\mathbf{X}$ of the child node $Y$. It remains to determine the mechanism set $\mathbf{M}$ and the surjection $\phi$, which both go hand-in-hand. The main purpose of SICI is to embed the assumption of ICI across the partitioned blocks of the parent set, rather than assuming this as a property of the singleton parents themselves. This objective is what the modeller should have in mind when determining the mechanism set and the surjection. Of course, it may be the case that the best surjection for embedding the ICI assumption is actually the bijection seen in the ICI model. In this case, the SICI model would reduce down to the ICI or PICI model, depending on the input source of stochasticity (noting that it would not make reasonable sense to model the upper relationships deterministically if $m=n$).

\begin{figure}[ht]
\centering
\scalebox{1.25}{
\begin{tikzpicture}
     \node[latent] (Y) {$Y$};%
     \node[obs,above=of Y,xshift=-2cm] (M1) {$M_{1}$}; %
     \node[obs,above=of Y,xshift=-0.8cm] (M2) {$M_{2}$}; %
     \node[obs,above=of Y,xshift=1cm] (Mm) {$M_{m}$}; %
     \node[latent,above=of M1,xshift=-1cm] (X1) {$X_1$};
     \node[latent,right=of X1,xshift=-0.8cm] (X2) {$X_2$};
     \node[latent,right=of X2,xshift=-0.8cm] (X3) {$X_3$};
     \node[latent,right=of X3,xshift=-0.8cm] (X4) {$X_4$};
     \node[latent,right=of X4,xshift=0.2cm] (Xn) {$X_n$};
     \node[below=of X2,yshift=0.8cm,xshift=-0.8cm] (f1){\small{$f_{(1)}$}};
     \node[below=of X4,yshift=0.8cm,xshift=-0.6cm] (f2){\small{$f_{(2)}$}};
     \node[below=of Xn,yshift=0.8cm,xshift=0.1cm] (fm){\small{$f_{(m)}$}};
      \node[below=of M1,yshift=0.3cm,xshift=1.1cm] (f){\small{$f$}};
     \path (M2) -- node[auto=false]{\ldots} (Mm);
     \path (X4) -- node[auto=false]{\ldots} (Xn);
    \coordinate[right of=X4,xshift=-0.35cm,yshift=-0.2cm] (d1);
    \coordinate[right of=d1,xshift=-0.4cm] (d2);
     \edge {M1,M2,Mm} {Y};
     \edge {X1,X2,X3} {M1};
     \edge{X4,d1} {M2};
     
     \edge {Xn,d2}{Mm};
\end{tikzpicture}
}
\caption{The SICI model structure with $\phi^{-1}(M_1)=\{X_1,X_2,X_3\}$, $\phi(X_4)=M_2$ and $\phi(X_n)=M_m$}
\label{SICIstructure}
\end{figure}
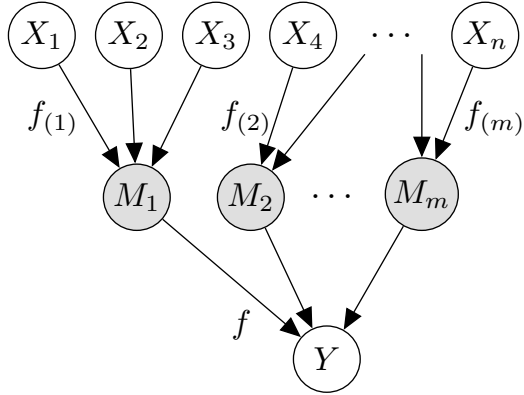

The mechanism nodes do not need to, and often don't, model explicit variables of the real-world system; a mechanism node, $M_i$, is instead simply defined by the subset of $\mathbf{X}$ that feeds into it, denoted $\phi^{-1}(M_i)=\mathbf{X}_{(i)}$, as well as by the CPT of $M_i\mid\mathbf{X}_{(i)}$, and does not represent a variable in its own right. Therefore, to arrive at the SICI model structure, we simply need to determine the partition of $\mathbf{X}$ that best embeds the ICI assumption across its blocks. Each block, $B_i$, formed of parents $\mathbf{X}_{(i)}$, of the partition is then directly connected to a common mechanism node, $M_i$, which does not require a precise semantic definition and whose indexing is insignificant. The relationships at each node still need quantifying, but we discuss this later. The general SICI model structure - for a given choice of surjection $\phi$ - is shown in Figure \ref{SICIstructure}.
\vspace{0.5cm}

Note that we can always ensure the SICI model is planar (i.e. a tree) by ordering the parents and mechanisms such that $\phi(X_1)=M_1$, $\phi(X_n)=M_m$ and:
\vspace{-0.1cm}
\begin{equation}\label{eq:planarity}
\begin{split}
    &\hspace{-1cm}\forall \; \; i \in \{1,\ldots,n-1\}, \; k \in \{1,\ldots,n-i\}, \\ &\phi(X_i)=\phi(X_{i+k}) \implies \phi(X_i)=\phi(X_{i+1})=\ldots=\phi(X_{i+k-1})=\phi(X_{i+k}).
\end{split}
\end{equation}

The surjective mapping between $\mathbf{X}$ and $\mathbf{M}$ in the SICI model slightly alters the conditional independence structure from that found in the ICI model. The SICI model structure is represented by the following conditional independence statements that can be verified through d-separation \cite{Pearl1988ProbReasoning,lauritzen1996graphical}:
\begin{align}
   & \hspace{-2cm} Y\indep \mathbf{X}\mid \mathbf{M} \label{eq:SICIindep1} \\ 
   & \hspace{-2cm} M_i\indep M_j\mid \mathbf{X} \label{eq:SICIindep2} \\
   & \hspace{-2cm} \mathbf{M}\indep \left(V\setminus\{\mathbf{X}\cup Y\}\right)\mid \{\mathbf{X}\cup Y\}\label{eq:SICIindep3} \\
       & \hspace{-2cm} M_i\indep \left(\mathbf{X}\setminus\mathbf{X}_{(i)}\right)\mid \mathbf{X}_{(i)}\label{eq:SICIindep4}
\end{align}
Note that while statements \ref{eq:SICIindep1}-\ref{eq:SICIindep3} remain unchanged from those of the ICI model structure (matching statements \ref{eq:ICIindep1}-\ref{eq:ICIindep3}), statement \ref{eq:SICIindep4} has been updated to reflect the surjective mapping onto the mechanism nodes, as opposed to the bijective mapping of the ICI model reflected previously in statement \ref{eq:ICIindep4}.

Once the surjection $\phi:\mathbf{X}\rightarrow\mathbf{M}$ has been determined, whether that be through data or expert judgement, it remains to quantify the relationships in the model. We now introduce the three variants of the SICI model that represent different areas of the local structure in which to input stochasticity.

\subsection{Lower-Stochastic SICI (LS-SICI)}
The first variant of SICI model is the lower-stochastic SICI (LS-SICI) model, featuring deterministic upper (parent-to-mechanism) relationships and a stochastic lower (mechanism-to-child) relationship. The functions $f_{(\cdot)}$ represent (compositions of) deterministic operators such as OR, AND, XOR and MAX, likely chosen through expert judgement (or even published domain literature) once the surjection $\phi$ has been determined. The parents that are grouped together in blocks will naturally be related to each other as they will share overlapping causal mechanisms affecting the child. Hence inferring sensible deterministic operators for each block through an expert elicitation workshop will not be such a challenge compared to directly eliciting a set of probabilistic judgements. These operators can be determined through natural language discussions with experts about the necessary requirements that each subset of parent nodes has in order to produce a given effect on the child. For example, it may only be necessary for either rainfall or wind speeds to be high to cause a reduction in honey bee foraging levels - hence the two variables may combine into a shared mechanism through an OR gate. If the experts are not satisfied with such a deterministic operator as an approximation to how the parents combine into their common causal mechanism, some stochasticity could be introduced, using these deterministic operators as a starting block from which to add noise. This process can be used to yield either the DS-SICI or US-SICI models seen below.

The function $f$ in the LS-SICI model represents the stochastic CPT of $Y|\mathbf{M}$. This CPT will typically have a smaller parameter space than that of the original CPT of $Y|\mathbf{X}$, helping facilitate efficient model parameterisation. We can additionally benefit from the embedding of the ICI assumption across the mechanism nodes $\mathbf{M}$ through the choice of surjection $\phi$. Because of this, the CPT of $Y|\mathbf{M}$ can be justifiably and faithfully approximated using quantitative CPT approximation techniques that rely on, or that can be enhanced by, the assumption of ICI. These techniques include ranked nodes approaches \cite{Fenton2007}, regression-based methods \cite{Rijmen2008,Alkhairy2020} and interpolation rules \cite{Cainsmethod,Hassall2019,Mascaro2022,Podofillini2015,Wisse2008}, as well as explicit ICI models such as noisy-OR \cite{Pearl1988ProbReasoning}, noisy-MAX \cite{Diez1993,Srinivas1993}, leaky noisy-OR \cite{Henrion1989} and the (leaky) intercausal cancellation model \cite{Woudenberg2015}, to name a few. Some of these methods have been previously reviewed to help guide the selection of a quantitative CPT approximation technique for a specific modelling problem \cite{Blomaard2025,Mkrtchyan2016}. These CPT approximation techniques combine with the reduced parameter space of the CPT of $Y|\mathbf{M}$ compared to that of $Y|\mathbf{X}$ (assuming $m<n$) to significantly improve the efficiency of the parameterisation of the local structure while maintaining faithfulness of the model.

Once the upper and lower relationships are defined, we can calculate the approximate CPT of $Y|\mathbf{X}$. Due to the deterministic nature of the upper relationships, multiple parent configurations $\mathbf{x}$ will deterministically map to the same configuration of mechanism values $\mathbf{m}$. Hence we will see several rows in the approximated CPT of $Y|\mathbf{X}$ featuring the same conditional probability distribution (CPD) over the child states. A common CPD will be shared by any row corresponding to a parent configuration $\mathbf{x}$ that deterministically obtains the mechanism configuration $\mathbf{m}$. To calculate the CPT of $Y|\mathbf{X}$, we therefore simply use the following approximation:
\begin{equation}\label{eq:LS-SICI}
    p(y|\mathbf{x})=p\left(y|\mathbf{m}=f(\mathbf{x})\right)
\end{equation}

The LS-SICI model provides significant parameter savings as the only probabilistic quantities required are those parameterising $Y|\mathbf{M}$. Full elicitation of the CPT $Y|\mathbf{M}$ will likely come at a reduced cost compared to direct elicitation of $Y|\mathbf{X}$. That said, the CPT approximation methods given above are well-suited to approximate this CPT as they make use of the embedding of the ICI property across the mechanism set $\mathbf{M}$. If using an approximation method to quantify this relationship, as few as $m$ parameters need to be determined, depending on the quantitative method chosen. The LS-SICI model shifts more focus onto qualitative elicitation than quantitative elicitation. In essence, it is possible to construct this model mainly through natural discussions with experts rather than eliciting probabilistic quantities. This makes the elicitation process for more intuitive and engaging for domain experts. The main question regarding suitability of the LS-SICI model is whether it is appropriate for rows in the CPT approximation of $Y|\mathbf{X}$ to share common CPDs. If it is feasible that there may be common rows in the true CPT, this method may be suitable to use at a low cost. However, if experts are not willing to accept that particular CPDs may be common across different parent configurations $\mathbf{x}$, it may be necessary to increase the flexibility of the local structure model through adding stochastic relationships into the upper half of the structure.

\subsection{Double-Stochastic SICI (DS-SICI)}
If more flexibility and expressiveness is desired in the LS-SICI model, the upper relationships can also be made stochastic. This yields the most general SICI model - the DS-SICI model - which models all nodes stochastically throughout the local structure. Being the most expressive SICI model, it features the largest parameter space of the SICI variants, and is therefore the most complex to parameterise. Figure \ref{SICIstructure} exactly represents the DS-SICI model as no node is drawn to be deterministic. Each combination function in the DS-SICI model represents a stochastic CPT.

In order to model the lower relationship, we can follow the same guidance as given for the LS-SICI model; the ICI property is embedded across the mechanism set $\mathbf{M}$, and we can therefore utilise quantitative CPT approximation methods referenced above and in Section \ref{BBNs} as a desirable alternative to direct elicitation to model the stochastic CPT of $Y|\mathbf{M}$.

We cannot, however, justifiably assume ICI to be a property across any subset of parents $\mathbf{X}_{(i)}$ sharing a common mechanism node $M_i$. Instead, we may determine these CPTs directly through data or expert judgement; the CPT $M_i|\mathbf{X}_{(i)}$ will likely be significantly smaller in the number of entries it bears compared to the original CPT $Y|\mathbf{X}$, hence direct elicitation of the mechanism CPTs will likely still yield a beneficial parameter saving. The difficulty with this arises through the fact that the mechanism nodes do not represent unique variables in their own right, hence probabilistic judgements about $M_i|\mathbf{X}_{(i)}$ are semantically ambiguous. Alternatively, it may be best to view each mechanism as a categorisation of the parents that feed into it, and to discuss with experts the requirements that are necessary for each categorisation of parents (i.e. for each mechanism) to produce an effect on the child. These requirements would naturally consist of logical operators, determined in the same way as for the LS-SICI model but with the addition of noise.

Given the stochastic nature of each relationship in the local model, we use the law of total probability and conditional independence statements \ref{eq:SICIindep1}-\ref{eq:SICIindep4} to calculate the approximated CPT of $Y|\mathbf{X}$ by:
\begin{equation}\label{eq:DS-SICI}
    p(y|\mathbf{x})=\sum_{\mathbf{m}}p(y|\mathbf{m},\mathbf{x})p(\mathbf{m}|\mathbf{x})=\sum_{\mathbf{m}}\left(p(y|\mathbf{m})\prod_{i=1}^mp(m_i|\mathbf{x}_{(i)})\right).
\end{equation}

The clear benefit of the DS-SICI structure is that it provides significantly more flexibility than the other SICI variants, allowing stochastic relationships to define every node in the model. This is therefore a model which is more likely to be accepted by domain experts as a reasonable approximation to of the real-world system. The increased resources needed to construct this model will likely lead to an increased level of satisfaction and faith in the model outputs - which is of high importance if the model is to be used for decision support. However, there is a much heavier focus on quantitative elicitation than qualitative elicitation when defining the DS-SICI structure; there is a high level of dependence on obtaining faithful probabilistic judgements from experts. This process needs to be carefully structured to ensure that the model inputs, and thus its outputs, are reliable. In addition, most of the quantitative elicitation concerns the modelling of the mechanism nodes which aren't representing variables in their own right. This is likely to make direct elicitation rather unintuitive. For this reason, it would often be wise to construct deterministic relationships to define the relationships going into the mechanism nodes. Therefore, we generally recommend constructing an LS-SICI model first before evaluating the need to add noise to these upper relationships to increase the flexibility of the model. Nonetheless, the DS-SICI model is generally easier to parameterise through expert judgement than the original local structure, and yet it is far more flexible than other local structure models.

\subsection{Upper-Stochastic SICI (US-SICI)}
If the DS-SICI model is unnecessarily complex for a given modelling problem, another alternative is to use the US-SICI model. Like the original ICI model, the US-SICI model features stochastic upper (parent-to-mechanism) relationships but a deterministic lower (mechanism-to-child) relationship. The US-SICI model would be shown in Figure \ref{SICIstructure} if the node $Y$ were drawn with two concentric circles, with only the function $f$ being deterministic.

For the lower relationship, we would again typically define $f$, representing $Y|\mathbf{M}$, in terms of (compositions of) deterministic logical operators over $\mathbf{M}$ such as OR, AND, XOR and MAX. This would likely be determined through natural language discussions with experts as part of an elicitation workshop, similar to how the upper relationships would be determined in the LS-SICI model. This would then leave the stochastic upper relationships, quantified through the CPT $M_i|\mathbf{X}_{(i)}$, to be determined. We cannot reasonably assume the ICI property to hold across any $\mathbf{X}_{(i)}$, hence the CPT may need to be elicited or otherwise determined directly, or through the addition of noise to some initially chosen composition of deterministic operators. This follows the exact same process as discussed for the DS-SICI model above.

Given that the lower relationships are now deterministic, in calculating $p(y|\mathbf{x})$, we now sum over configurations of mechanism values that lead to the event $Y=y$ deterministically - analogously to that seen in Equation \ref{ICIdef}. We hence amend Equation \ref{eq:DS-SICI} to account for this to obtain:
\begin{equation}\label{eq:US-SICI}
    p(y|\mathbf{x})=\sum_{\mathbf{m}}p(y|\mathbf{m},\mathbf{x})p(\mathbf{m}|\mathbf{x})=\sum_{\mathbf{m}\mid f(\mathbf{m})=y}\prod_{i=1}^mp(m_i|\mathbf{x}_{(i)}).
\end{equation}


The US-SICI model is a useful alternative to the LS-SICI model if each row of the approximate CPT having its own, unique CPD is a desired property of the model. The model may in some cases require fewer quantitative parameters than the LS-SICI model due to it featuring a larger number of smaller CPTs. This is likely to occur when the partition of the parent set is relatively fine, featuring small blocks and a larger number of mechanism nodes $m$. If this parameter saving over the LS-SICI model is notable, it may be worth using the US-SICI model. When this parameter saving is less significant, even when the LS-SICI model features a slightly larger number of parameters, the LS-SICI model may be a better choice as it enables the justified use of quantitative CPT approximation methods through the embedding of the ICI property over the mechanism set. The US-SICI model does not make use of this property due to its deterministic modelling of $Y|\mathbf{M}$, hence its parameters cannot be determined so efficiently. Further, parameterising the US-SICI model may be less intuitive than parameterising the LS-SICI model as the mechanism nodes do not often correspond the explicit real-world variables in their own right. Some modelling applications may render the US-SICI model the best approach, though the LS-SICI and DS-SICI models would be preferred in many cases.

\subsection{Examples}

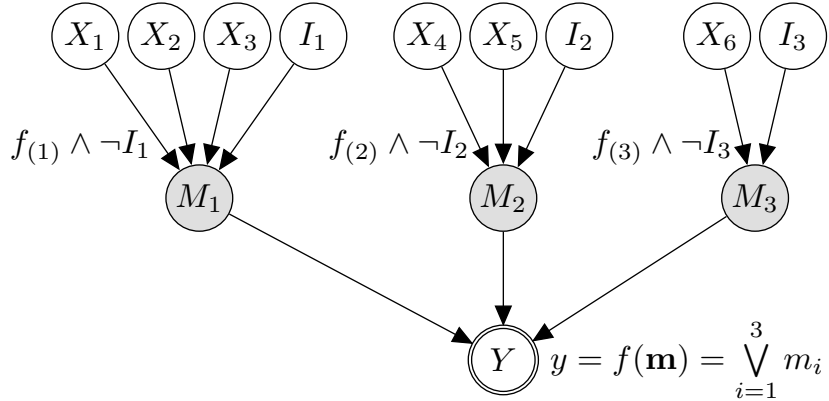
\begin{figure}[ht]
\centering
\scalebox{1.25}{
\begin{tikzpicture}
     \node[latent,style = double] (Y) {$Y$};%
     \node[obs,above=of Y,xshift=-3.2cm] (M1) {$M_{1}$}; %
     \node[obs,above=of Y,xshift=0cm] (M2) {$M_{2}$}; %
     \node[obs,above=of Y,xshift=2.65cm] (Mm) {$M_{3}$}; %
     \node[latent,above=of M1,xshift=-1.2cm] (X1) {$X_1$};
     \node[latent,above=of M1,xshift=-0.4cm] (X2) {$X_2$};
     \node[latent,above=of M1,xshift=0.4cm] (X3) {$X_3$};
     \node[latent,above=of M1,xshift=1.2cm] (I1) {$I_1$};
     \node[latent,above=of M2,xshift=-0.8cm] (X4) {$X_4$};
     \node[latent,above=of M2,xshift=0cm] (X5) {$X_5$};
     \node[latent,above=of M2,xshift=0.8cm] (I2) {$I_2$};
     \node[latent,above=of Mm,xshift=-0.4cm] (Xn) {$X_6$};
     \node[latent,above=of Mm,xshift=0.4cm] (I3) {$I_3$};
     \node[below=of X2,yshift=0.5cm,xshift=-0.85cm] (f1){\small{$f_{(1)}\land\lnot I_1$}};
     \node[below=of X4,yshift=0.5cm,xshift=-0.3cm] (f2){\small{$f_{(2)}\land\lnot I_2$}};
     \node[below=of Xn,yshift=0.5cm,xshift=-0.58cm] (fm){\small{$f_{(3)}\land\lnot I_3$}};
    \coordinate[right of=X4,xshift=-0.35cm,yshift=-0.2cm] (d1);
    \coordinate[right of=d1,xshift=-0.4cm] (d2);
     \edge {M1,M2,Mm} {Y};
     \edge {X1,X2,X3,I1} {M1};
     \edge{X4,X5,I2} {M2};
     
     \edge {Xn,I3}{Mm};
     \node[right=of Y,yshift=0cm,xshift=-1cm] (f){\small{$y=f(\mathbf{m})=\bigvee\limits_{i=1}^{3}m_i$}};;
\end{tikzpicture}
}
\caption{The surjective noisy-OR Model - a member of the SICI model family}
\label{SICIGNoisyOR}
\end{figure}

We now introduce an example of a causal interaction model that is a member of the SICI model family. This is a generalisation of the noisy-OR model \citep{Pearl1988ProbReasoning} which we name the surjective noisy-OR model. In this model on a set of binary nodes, the parent set is partitioned into blocks that share a common inhibitor variable; parents in the same block feed into a shared mechanism node, say, $M_i$, just as described above for the general SICI model. However, just like the standard noisy-OR model, we also introduce an inhibitor node $I_i$ which also feeds into the mechanism node $M_i$ for each $i=1,\ldots,m$. The parent-to-mechanism combination function for mechanism $M_i$ is now denoted $f_{(i)}\land\lnot I_i$, where $f_{(i)}$, as before, describes how the parents in the block collectively influence the child through their common causal mechanism. The activation of the mechanism node $M_i$ now further depends on this causal mechanism not being inhibited by $I_i$. The mechanism-to-child combination function is simply the deterministic OR function over the mechanism nodes. This model satisfies the definition of the SICI model and, in particular, is a US-SICI model as the inhibitor node is defined stochastically (as may be the combination functions $f_{(i)}$). The surjective noisy-OR model is shown in Figure \ref{SICIGNoisyOR} for a given mapping $\phi$ on 6 parents. Note that, while the functions $f_{(i)}$ need to be determined, the number of quantitative parameters to be determined has fallen from 6 for the standard noisy-OR model to 3 for the surjective noisy-OR model - in general giving a saving of $n-m$ parameters.

\vspace{0.5cm}

We will further demonstrate the use of LS-SICI and DS-SICI models through exploring the application of Hassall's algorithm \cite{Hassall2019} within the SICI framework. Hassall's algorithm is a simple CPT approximation technique based on linear interpolation over the whole interval $[0,1]$. It involves the elicitation of weights for each parent, and assumes that states are ordered and equally spaced. For simplicity, we will apply Hassall's algorithm to the case of a local model composed entirely of binary variables, though Hassall's algorithm extends to multi-state parents and children. In the binary case, once we have elicited relative weights $w_i$ for each parent $X_i$, we obtain the following approximation \cite{Hassall2019}:
\begin{align}\label{eq:HassallBinary}
    \mathbb{P}(Y=1|\mathbf{X}=\mathbf{x})=\frac{\sum\limits_{i=1}^nw_iX_i}{\sum\limits_{i=1}^nw_i}.
\end{align}

We will now demonstrate that this approximation is actually a probabilistic ICI (PICI) model. Suppose we introduce a set of latent mechanisms $\mathbf{M}$ such that $M_i|X_i\sim\text{Bernoulli}(w_ix_i)$. Further, suppose that $Y|\mathbf{M}\sim\text{Bernoulli}\left(\frac{\sum_{i=1}^nm_i}{\sum_{i=1}^nw_i}\right)$. Importantly, for this PICI set-up, the weights must be normalised such that they sum to $n$ (i.e. $\sum_{i=1}^nw_i=n$) in order for $\mathbb{P}(Y=1|\mathbf{X})$ to be mapped onto the whole interval $[0,1]$ as seen in Hassall's algorithm. When using Hassall's algorithm in practice, no intermediate nodes would need to be introduced, thus this specific normalisation requirement would not be needed. The above set-up corresponds to a PICI structure with the following quantification, mirroring Equation \ref{eq:PICI}:
\begin{align}
    \mathbb{P}(Y=1|\mathbf{x})=\sum_\mathbf{m}\left[\left(\frac{\sum_{i=1}^nm_i}{\sum_{i=1}^nw_i}\right)\prod_{i=1}^n(w_ix_i)^{m_i}(1-w_ix_i)^{1-m_i}\right].
\end{align}
We can further note the following:
\begin{align}
    \mathbb{P}(Y=1|\mathbf{x})=&\sum_\mathbf{m}\left[\left(\frac{\sum_{i=1}^{n}m_i}{\sum_{i=1}^{n}w_i}\right)\prod_{i=1}^n(w_ix_i)^{m_i}(1-w_ix_i)^{1-m_i}\right] \nonumber \\
    =&\mathbb{E}_{\mathbf{M|x}}\left[\frac{\sum_{i=1}^{n}M_i}{\sum_{i=1}^{n}w_i}\right] \nonumber\\
    =&\frac{1}{\sum_{i=1}^{n}w_i}\sum_{i=1}^n\mathbb{E}_{\mathbf{M}_i|\mathbf{x}_i}\left[M_i\right]=\frac{\sum_{i=1}^nw_ix_i}{\sum_{i=1}^nw_i}
\end{align}
Hence we have that:
\begin{align}
    p(y|\mathbf{x})=\frac{\sum_{i=1}^nw_ix_i}{\sum_{i=1}^nw_i} \nonumber&=\sum_m\left[\left(\frac{\sum_{i=1}^nm_i}{\sum_{i=1}^nw_i}\right)\prod_{i=1}^n(w_ix_i)^{m_i}(1-w_ix_i)^{1-m_i}\right] \nonumber\\
    &=\sum_\mathbf{m}\left[p(y|\mathbf{m})\prod_{i=1}^np(m_i|x_i)\right]
\end{align}
Therefore, Hassall's algorithm on a binary set of nodes can be expressed as a PICI model, through the introduction of the latent mechanism set $\mathbf{M}$, in which $M_i|X_i\sim\text{Bernoulli}(w_ix_i)$ and $Y|\mathbf{M}\sim\text{Bernoulli}\left(\frac{\sum_{i=1}^{n}m_i}{\sum_{i=1}^{n}w_i}\right)$.

Hassall's algorithm - as a CPT approximation technique that implicitly assumes ICI among parent nodes - could be used to model the lower mechanism-to-child nodes in a DS-SICI or LS-SICI model. This would be making use of the objective to embed the ICI property across the set of mechanisms through the choice of surjection $\phi$. We can do this whether the upper relationships are stochastic or deterministic. 

Suppose we have three mechanism nodes (i.e. $m=3$), all of which are binary, as is the child node $Y$. In order to apply Hassall's algorithm, we would need to determine - most likely through expert judgement - relative weights $w_i$ for each of the mechanism nodes $M_i$. This can be performed using ideas referenced in Section \ref{BBNs}. It will often be possible, and indeed quite useful, to view each mechanism node as a particular categorisation of the parent nodes and to score the relative influence of each category. Keeping generality, suppose we have weights $w_1$, $w_2$ and $w_3$ for $M_1$, $M_2$ and $M_3$ respectively. As each node is binary, we use Equation \ref{eq:HassallBinary} to obtain the generic CPT of $Y|\mathbf{M}$ shown in Table \ref{tab:HassallCPT}.

\begin{table}[h!]
\centering
\caption{Generic CPT of $Y|\mathbf{M}$ using Hassall's algorithm}
\label{tab:HassallCPT}
\renewcommand{\arraystretch}{1.032}
\begin{tabular}{|c|c|c|c|c|}
\hline
\( M_1 \) & \( M_2 \) & \( M_3 \) & \( \mathbb{P}(Y=1 \mid \mathbf{M}=\mathbf{m}) \) & \( \mathbb{P}(Y=0 \mid \mathbf{M}=\mathbf{m}) \) \\
\hline
0 & 0 & 0 & \( 0 \) & \( 1 \) \\
1 & 0 & 0 & \( \frac{1}{W}(w_1) \) & \( 1 - \frac{1}{W}(w_1) \) \\
0 & 1 & 0 & \( \frac{1}{W}(w_2) \) & \( 1 - \frac{1}{W}(w_2) \) \\
0 & 0 & 1 & \( \frac{1}{W}(w_3) \) & \( 1 - \frac{1}{W}(w_3) \) \\
1 & 1 & 0 & \( \frac{1}{W}(w_1 + w_2) \) & \( 1 - \frac{1}{W}(w_1 + w_2) \) \\
1 & 0 & 1 & \( \frac{1}{W}(w_1 + w_3) \) & \( 1 - \frac{1}{W}(w_1 + w_3) \) \\
0 & 1 & 1 & \( \frac{1}{W}(w_2 + w_3) \) & \( 1 - \frac{1}{W}(w_2 + w_3) \) \\
1 & 1 & 1 & \( 1 \) & \( 0 \) \\
\hline
\end{tabular}
\end{table}
\vspace{0.25cm}

This CPT provides us the eight CPDs corresponding to each configuration of the mechanism set $\mathbf{M}$. These CPDs are used to then replace the $p(y|\mathbf{M})$ term in the approximation of $p(y|\mathbf{x})$ in either Equation \ref{eq:DS-SICI} for the DS-SICI model or Equation \ref{eq:LS-SICI} for the LS-SICI model. If using the LS-SICI model, due to the deterministic upper relationships it features, the rows in the CPT of $Y|\mathbf{X}$ will feature the CPDs from Table \ref{tab:HassallCPT} without modification. Each CPD from Table \ref{tab:HassallCPT} would appear in any row representing a configuration of parent values that deterministically maps to the corresponding configuration of mechanism values from which the CPD stems. If using the DS-SICI model, each CPD in Table \ref{tab:HassallCPT} would form part of 
weighted sum over all configurations of mechanism values, weighted by the probability of each such configuration occurring according to the stochastic upper relationships. In either case, Hassall's algorithm is just one CPT approximation method implicitly assuming the ICI property that can then be used to model this lower mechanism-to-child relationship.

\section{Discussion} \label{Discussion}

In this paper, we introduce the surjective independence of causal influences (SICI) model as a generalisation of the ICI model. We specifically introduce three variants of the SICI model - the DS-SICI, US-SICI and LS-SICI models, granting the modeller full flexibility regarding the input source of stochasticity in the model. We additionally provide some guidance about how to choose between these models through addressing the benefits and limitations of each variant. 

Each of the SICI models can be elicited primarily through natural language conversations with experts, with less dependence than direct elicitation on obtaining probabilistic judgements that is so burdensome to experts. It naturally combines with existing quantitative CPT approximation methods that facilitate efficient quantitative elicitation of ICI structures - here composed of a partition of the parent set. The SICI model is significantly less restrictive than the ICI model through its construction explicitly allowing flexibility in the source of stochasticity, as well as through allowing interactions between parents. The interactions can be embedded through simple compositions of deterministic logical operators, or more complex, possibly stochastic, relationships, if desired. This can be achieved with few, if any, quantitative expert judgements. The CPT of $Y|\mathbf{M}$ in the SICI model can be \textit{justifiably} approximated through existing quantitative CPT approximation methods alluded to in Section \ref{BBNs} as a result of the ICI property being embedded across the mechanism set through the choice of surjection $\phi$. This is demonstrated in Section \ref{SICI} through the application of Hassall's algorithm \cite{Hassall2019} within the SICI model. The use of such CPT approximation methods produces significant parameter savings to the Bayesian network parameterisation process, reducing the elicitation burden, if using expert judgement, or reducing data requirements if using data-driven parameter learning techniques. 

The SICI model does introduce additional complexity to the qualitative elicitation stage, though this is far less burdensome for experts than a complex quantitative elicitation process. We therefore argue that this shift in complexity comes with a net benefit considering just the elicitation burden. As a result, the SICI methodology facilitates quicker achievement of the modelling objectives while significantly reducing client resource requirements. In addition to the ability to model interactions, the SICI model can handle large parent sets more efficiently than the ICI model, eliminating the practical need to compromise on small parent sets. These factors ensure that the SICI model enables more faithful modelling than the ICI model, giving the client crucial faith in the model outputs.

The underlying assumption of ICI is not one that we have deemed satisfactory in many practical modelling domains. The SICI methodology satisfies the clear need to generalise the ICI model to weaken this assumption, enabling faithful BN modelling with minimal client resource requirements. Our next steps in the development of the SICI methodology are to develop a complete BN model for an existing research application of ours fully utilising the SICI methodology; to compare this model to an existing BN model for the same application that does not utilise SICI; to explore the performance of each variant of the SICI model; and to develop a complete framework for the construction of elicited BN models that fully incorporates this SICI methodology with adapted existing quantitative elicitation methodologies. SICI is, of course, not the only method that facilitates efficient CPT approximation and BN parameterisation, and we will continue to explore and report on other methods going forward.

\end{document}